\documentstyle[12pt,epsf,epsfig]{aipproc}

\begin{document}
\sloppy         
\title{\LARGE\bf Tests of a fiber detector concept for high 
rate particle tracking}

\author{E.C. Aschenauer, J. B\"ahr, V. Gapienko\footnote{on leave from
IHEP Protvino, Russia}, B. Hoffmann\footnote{now at Esser Networks
GmbH, Berlin},\\
H. L\"udecke,
A. Menchikov\footnote{on leave from JINR Dubna},
C. Mertens\footnote{Summerstudent from University of Clausthal
Zellerfeld},\\ R. Nahnhauer, R. Shanidze\footnote{on leave from
High Energy Physics Institute, Tbilisi State University}}

\address{DESY-Zeuthen, 15738 Zeuthen, Germany}

\maketitle

\begin{abstract}
         A fiber detector concept is suggested allowing to registrate
         particles within less than 100 nsec with a space point
         precision of about 100 $\mu$m at low occuppancy. The fibers
         should be radiation hard for 1 Mrad/year.
           Corresponding prototypes have been build and tested at a 3 GeV
         electron beam at DESY. Preliminary results of these tests
         indicate that the design goal for the detector is reached.  
\end{abstract}

\section*{Introduction}
The advantageous use of fiber detectors for particle tracking has
been demonstrated for very different conditions e.g. in the
UA2-Experiment \cite{lit1}, CHORUS \cite{lit2}, for D0 \cite{lit3} 
and the H1-Forward Proton Spectrometer \cite{lit4}. Due to the 
different experimental situation in this applications 
three types of optoelectronic read out techniques are
applied -- Image Intensifier plus CCD's,
Visible Light Photon Counters and Position Sensitive
Photomultipliers. However, all have in common
that the precision of space point measurements is given by hits of
overlapping fibers of several staggered fiber layers. For high rate
experiments demanding online tracking of several hundred particles per
100 nsec bunch crossing such a concept may not work due to too high
occupancy of single fiber channels.

We propose in the following to use overlapping fiber roads reading
out several thin scintillating fibers with one clear optical
fiber. The demands and the solutions presented below match to a
possible application of the detector as the inner tracker in the
HERA-B project at DESY \cite{lit5}. Similar ideas have been used by
others \cite{lit6} to build a fiber detector for the DIRAC experiment at CERN.

\section*{Detector principle}
The fiber detector under discussion is aimed to detect throughgoing
particles with more than 90 \% efficiency within less than 100 nsec and a
precision of better than 100 $\mu$m. The fibers should not change their
characteristics significantly after an irradiation of 1 -- 2 Mrad. The
sensitive detector part should have a size of ~25 x 25 cm$^2$. The
scintillating fibers should be coupled to clear optical fibers of
about 3m length guiding the light to photosensors placed outside the
experimental area.

\begin{figure}[t] 
\begin{minipage}[b]{7cm}
\epsfig{file=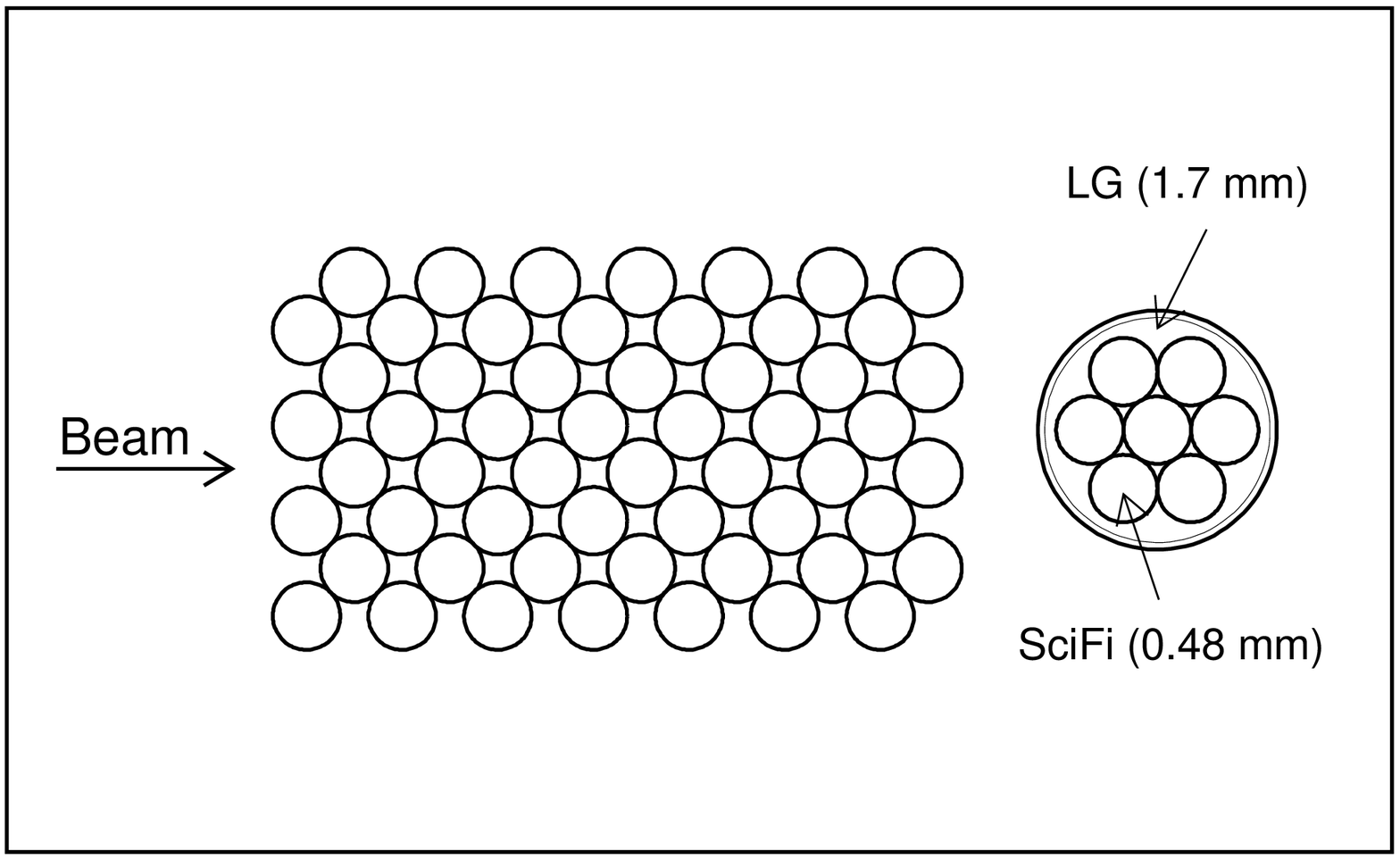,height=7cm}

\vspace{-0.7cm}
\caption{Schematic view of the proposed fiber detector 
 cross section and coupling principle (LG: light guide fiber)}
\label{bild1}
\end{minipage}
\hfill
\begin{minipage}[b]{7cm}
\epsfig{file=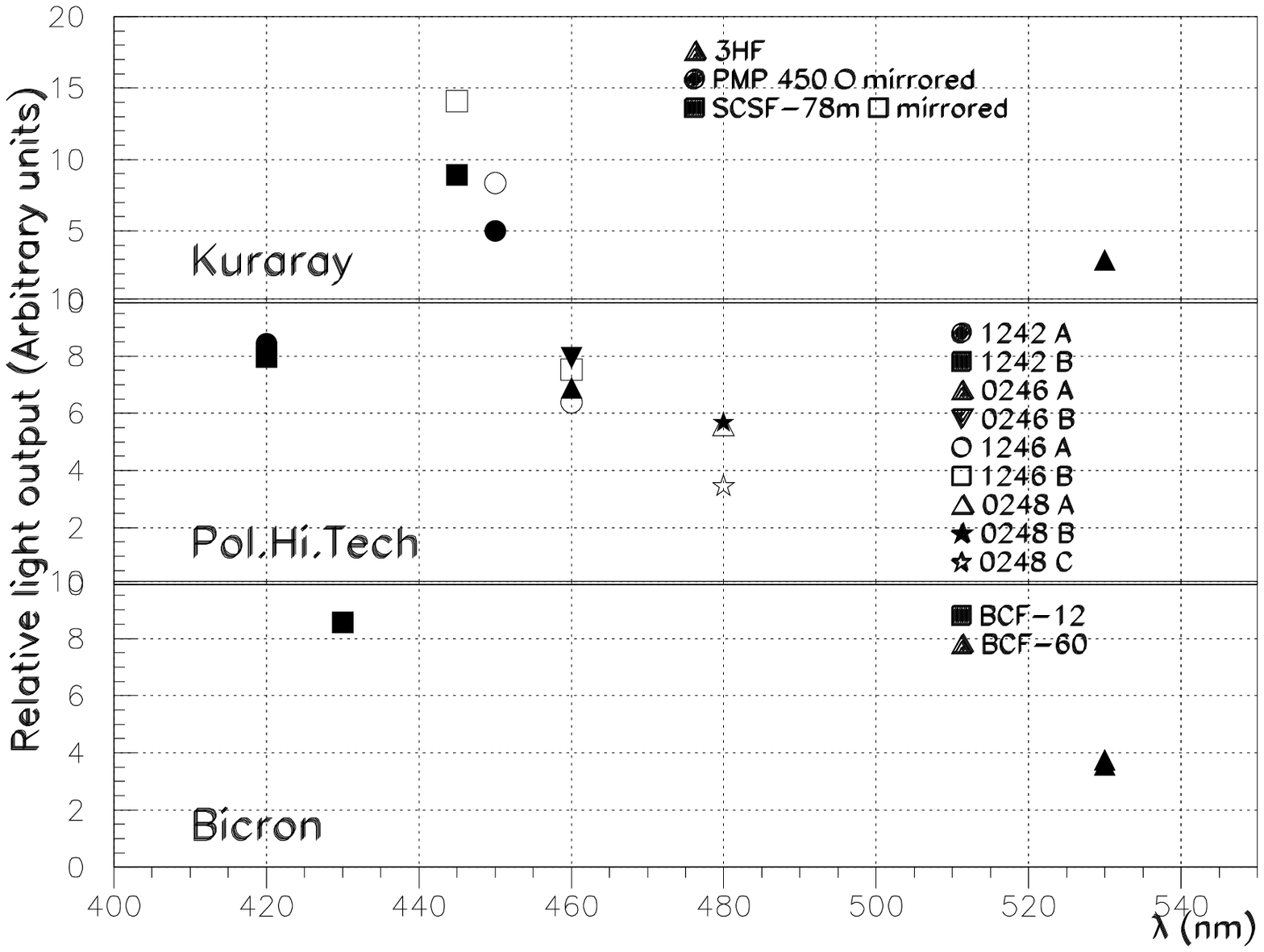,height=7cm}
\caption{Light output from 500 $\mu$m diameter fibers for
several fiber materials of three producers}
\label{bild2}
\end{minipage}
\end{figure}

It is assumed that most particles of interest hit the detector
perpendicular, i.e. with angles less than five degrees with respect to
the beam axis. In this case low occupancy and high light yield  are
guaranteed by using overlapping fiber roads like schematically drawn in
fig. \ref{bild1}. One fiber road consists of several thin scintillating fibers
arranged precisely behind each other and coupled to one thick light
guide fiber. The scintillating fiber diameter determines the space
point resolution of the detector. The number of fibers per road is
fixed by the scattering angle of particles and the allowed
amount of multiple scattering. It will also influence the factor of
background suppression for tracks with larger inclination or
curvature. The pitch between fiber roads is defined by demanding a
homogeneous amount of fiber material across the detector width.

Keeping in mind the conditions at HERA-B, we made the following
choices:

\begin{center}
\begin{tabular}{lrllrl}
 $\Phi_{fib}$ &=&480$\mu$m~~~~~~   & N$_{fib/road}$ &=&  7 \\
L$_{fib}$     &=&30 cm    & p$_{road}$     &=& 340 $\mu$m\\
$\Phi_{lg}$   &=&1.7 mm     & N$_{road}$   &=&   640 \\  
L$_{lg}$      &=&300 cm    & W$_{det}$     &=&  217.6 mm 
\end{tabular}
\end{center}

\noindent
with $\Phi$ and L: diameter and length of scintillating and light guide
fibers, N$_{fib/road}$: number of fibers per road, p$_{road}$: distance between
neighboured road centers, N$_{road}$: number of roads per detector, W$_{det}$:
detector width.

The light guide fibers are read out with the new
Hamamatsu\footnote{Hamamatsu Photonics K.K., Electron tube division,
    314--5, Shimokanzo, Tokyooka Village. Iwatagun, Shizuoka--ken,
    Japan} 64
channel PSPM R5900--M64 with a pixel size of 2 x 2 mm$^2$ \cite{lit7}. 
To diminish optical cross talk the thickness of the entrance 
window of the device was decreased to 0.8 mm.

The coupling between scintillating and light guide fibers is done by
loose plastic connectors. The light guides are coupled to the PSPM
using a plastic mask fitting the corresponding pixel pattern.

\section*{Material studies}
Double clad fibers  of three different producers\footnote{BICRON,
  12345 Kinsman Road, Newbury, Ohio, USA} \footnote{Pol. Hi. Tech.,
  s.r.l., S.P. Turanense, 67061 Carsoli(AQ), Italy} \footnote{KURARAY
  Co. LTD., Nikonbashi, Chuo-ku, Tokyo 103, Japan} were tested
concerning light output, light attenuation and radiation hardness for
several fiber diameters and wavelengths of the emitted light. Details of
these measurements are given in \cite{lit8}. A few results are summarized below.

The light output of fibers of 500 $\mu$m diameter is shown
in fig. \ref{bild2}. Generally it can be seen, that the light yield decreases
with increasing scintillator emission wavelength because the PM sensitivity
curve is not unfolded. There is no remarkable difference between the
best materials of the three producers. A mirror at the end of the
fiber increases the light output by a factor 1.7.

Several tests were performed to couple scintillating and light guide
fibers. Finally the coupling efficiency became better than 95 \%,
independent of the medium between both parts (air,glue,grease).

\begin{figure}[b] 
\begin{minipage}[b]{7cm}
\epsfig{file=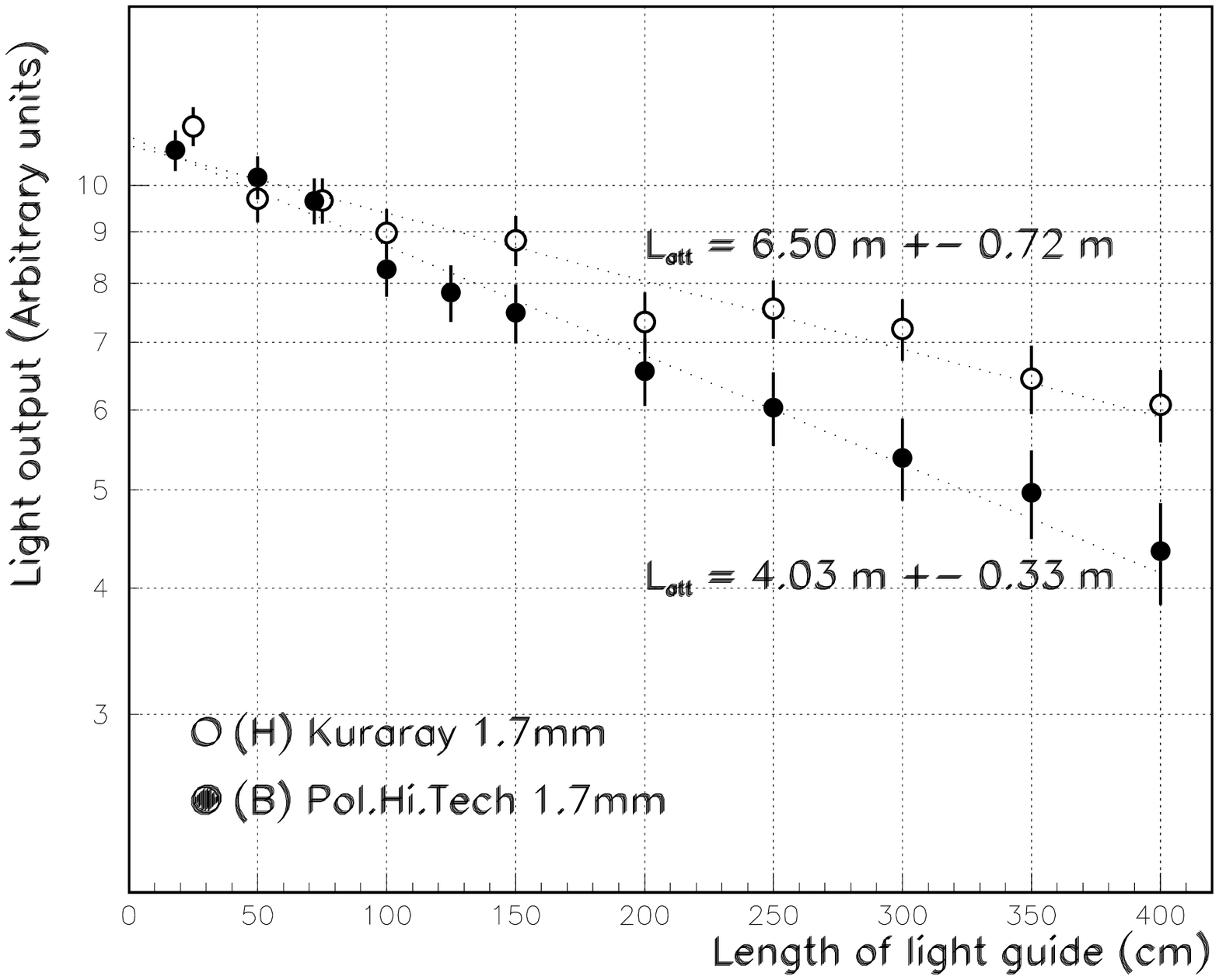,height=7cm}

\vspace{1.2cm}
\caption{Light attenuation in clear fibers of 1.7 mm diameter
produced by Kuraray and Pol.Hi.Tech.}
\label{bild3}
\end{minipage}
\hfill
\begin{minipage}[b]{7cm}
\epsfig{file=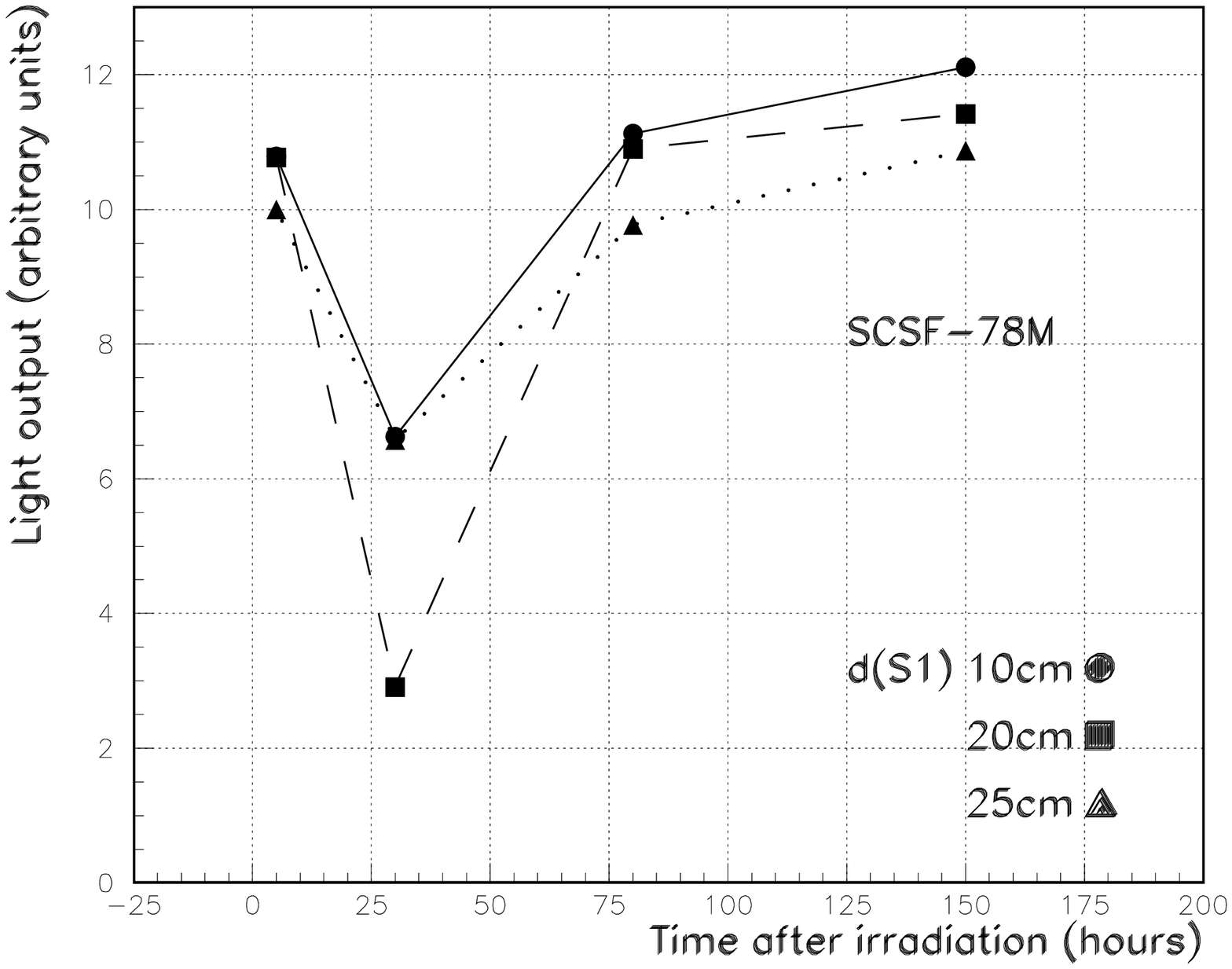,height=7cm}
\caption{Time evolution of light output for KURARAY SCSF 78M
fibers irradiated with 0.2 and 1.0 Mrad at 10 and 20 cm
respectively. The solid, dashed and dotted curves
correspond to measurements with a source placed
at 10,20 and 25 cm.}
\label{bild4}
\end{minipage}
\end{figure}

The light attenuation of clear fibers was measured coupling them to
single scintillating fiber roads excited by a Ruthenium source. The
clear fibers were cutted back piece by piece to the length under
investigation. Results for two producers are given in fig. \ref{bild3}.

Radiation hardness tests of fibers were made using an intense 70 MeV
proton beam at the Hahn--Meitner--Institute Berlin. 1 Mrad radiation was
deposited within a few minutes. For all materials investigated we
observed a damage of the scintillator and the transparency of the
fiber which was followed by a long time recovery of up to 600 h. An
example is shown in fig \ref{bild4}. More detailed studies using glued and
nonglued fibers and irradiate them in air and nitrogen atmosphere are
still ongoing.

Summarizing all results of our material studies we decided to use
the KURARAY fibers SCSF-78M with a diameter of 480 $\mu$m for the
scintillating part of our detector prototypes. For clear fibers still
two choices seem to be possible: 1.7 mm fibers from KURARAY or
Pol. Hi. Tech.. 

\section*{Detector production}
Using winding technology as developed for the CHORUS experiment \cite{lit9}
we built a detector production chain at our institute. A drum of 80 cm
diameter allows to produce five detectors at once. The production time
for winding one drum is about 14 h. Sticking the fibers to the
connector holes is still done by hand and rather time consuming. A part
of the polished end face of one of our detectors is shown in fig. 5.

Two other detector prototypes are ordered from
industry. GMS--Berlin\footnote{GMS - Gesellschaft f\"ur Mess- und
 Systemtechnik mbH, Rudower Chaussee 5, 12489 Berlin, Germany}
followed a technology proposed by the university of Heidelberg \cite{lit10}
mounting single layers on top of each other using epoxy glue. Each
layer is prepared on a v-grooved vacuum table. One layer per day can
be produced in this case. The connector is here also added by hand. 
The production technology used by KURARAY is unknown to us.

To get the precision of the detector geometry quantified we
measured the coordinates of all fibers of the polished end face of the
three detectors. In fig. 6 the deviation from the ideal position is
given per fiber road. Some stronger local effects are
visible. Averaging these results characteristic accuracies of
20 $\mu$m, 50 $\mu$m and 10 $\mu$m are calculated for the Zeuthen, GMS and KURARAY
detectors respectively.

\section*{Testrun results}
Two testruns were performed to measure the properties of the
produced fiber detectors in a 3 GeV electron beam at DESY. The setup
used in both cases was very similar and is schematically drawn in
fig. \ref{bild7}. Four silicon microstrip detectors are used together with two
fiber reference detectors and an external trigger system to predict
the position of a throughgoing particle at the detector to be
tested. A precision of 50 $\mu$m and 80 $\mu$m was reached for that prediction
using the geometrical arrangements of testrun 1 and 2. The fiber
detector signals were registrated after 3m of light guide in the first
case using a 16 channel PM R5900--M16 read out with a charge sensitive
ADC. In the later run the 64 channel R5900--M64 was used and the
signals were transfered via a special multiplexer to a flash ADC.

In April 1997 first small eight road detectors were investigated to
measure the light profile across a fiber road. The result is shown in
fig. 8. The data can be described simply by taking into account fiber
geometry seen by a throughgoing particle. They allow to calculate the
detector efficiency for any particular pitch between the fiber roads.

During the testrun in October 1997 the three full size detector
prototypes described in section 4 were investigated in detail. Up to
now only preliminary results are derived from about 4 Gbyte of data.

A relation of 0.9/1.0/0.8 was found for the average
light output of the Zeuthen, GMS and KURARAY detectors. It seems to be
due to the different quality of the end face polishing rather than to
the mechanical detector precision.

The detector efficiency and resolution is dependent on the hit
selection method used. With a maximum amplitude search for all PSPM
pixels we calculated rough values of 97 $\pm$ 3 \% for the efficiency and
about 140 $\mu$m for the resolution of the three detectors. (see also figs.
9 and 10). Taking into account the finite resolution of our track
prediction of 80 $\mu$m and the total mechanical alignment not better than
50 $\mu$m this points to a fiber detector resolution of better than 100 $\mu$m.

Work is in progress to qualify these results. In addition the
detector noise has still to be studied in detail. Optical and
electrical cross talk will influence the choice of cuts and the 
hit selection methods and in this way also efficiencies and resolution. 

\section*{Summary}
Three fiber detector prototypes have been tested. They are made out of 640
overlapping roads of seven 480 $\mu$m diameter fibers coupled to 1.7 mm
diameter light guides of 3 m length read out with 64 channel
photomultipliers.
  For all three detectors a preliminary analysis gives an efficiency
of about 97 \% and a resolution of about 100 $\mu$m. These results
together with radiation hardness studies of the used fiber material
seem to make it possible to use a corresponding detector in a high
rate experiment like HERA-B. In such case special care has to be taken
to keep noise from optical and electrical cross talk at an acceptable level.

\vspace*{0.5cm}

\noindent
{\bf Acknowledgement}\\
Part of this work was done in close collaboration with groups from the
universities of Heidelberg and Siegen. We want to thank our colleagues
for their good cooperation and many fruitful discussions.

The fiber irradiation tests were possible only due to the kind support
of the Hahn-Meitner-Institute Berlin. We are deeply indebted to the
ISL accelerator team and want to thank in particular
Dr. D. Fink, Dr. K. Maier and Dr. M. M\"uller from HMI and Prof. Klose from
GMS for a lot of practical help.

We acknowledge the benefit from the DESY II accelerator crew and
the test area maintainance group.

\newpage

\clearpage

\newpage

\begin{figure}[t] 
\begin{minipage}[b]{7cm}
\epsfig{file=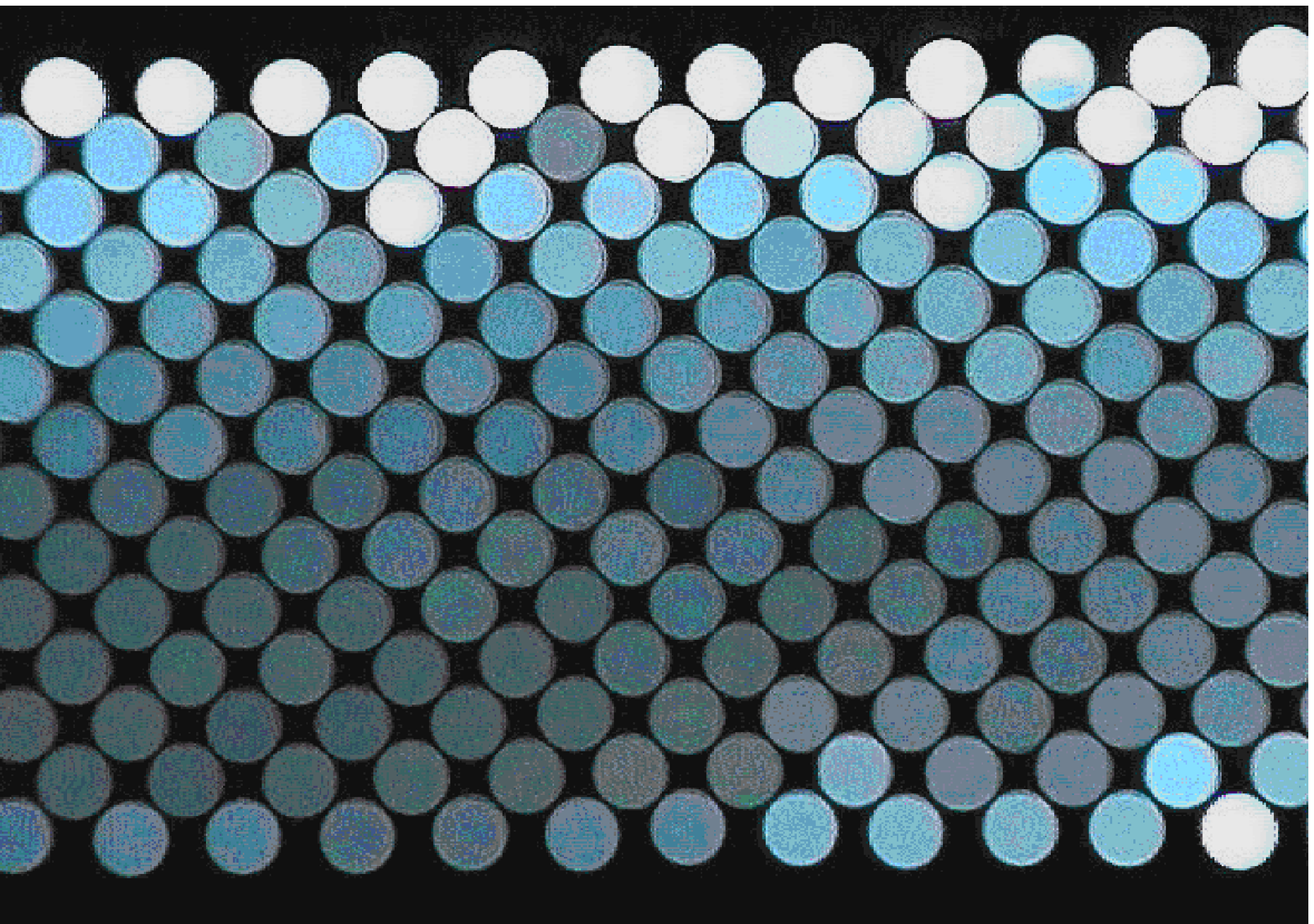,width=7cm}

\vspace{1.3cm}
\caption{Photograph of part of the polished end face of a Zeuthen
prototype detector.} 
\label{bild5}
\end{minipage}
\hfill
\begin{minipage}[b]{7cm}
\epsfig{file=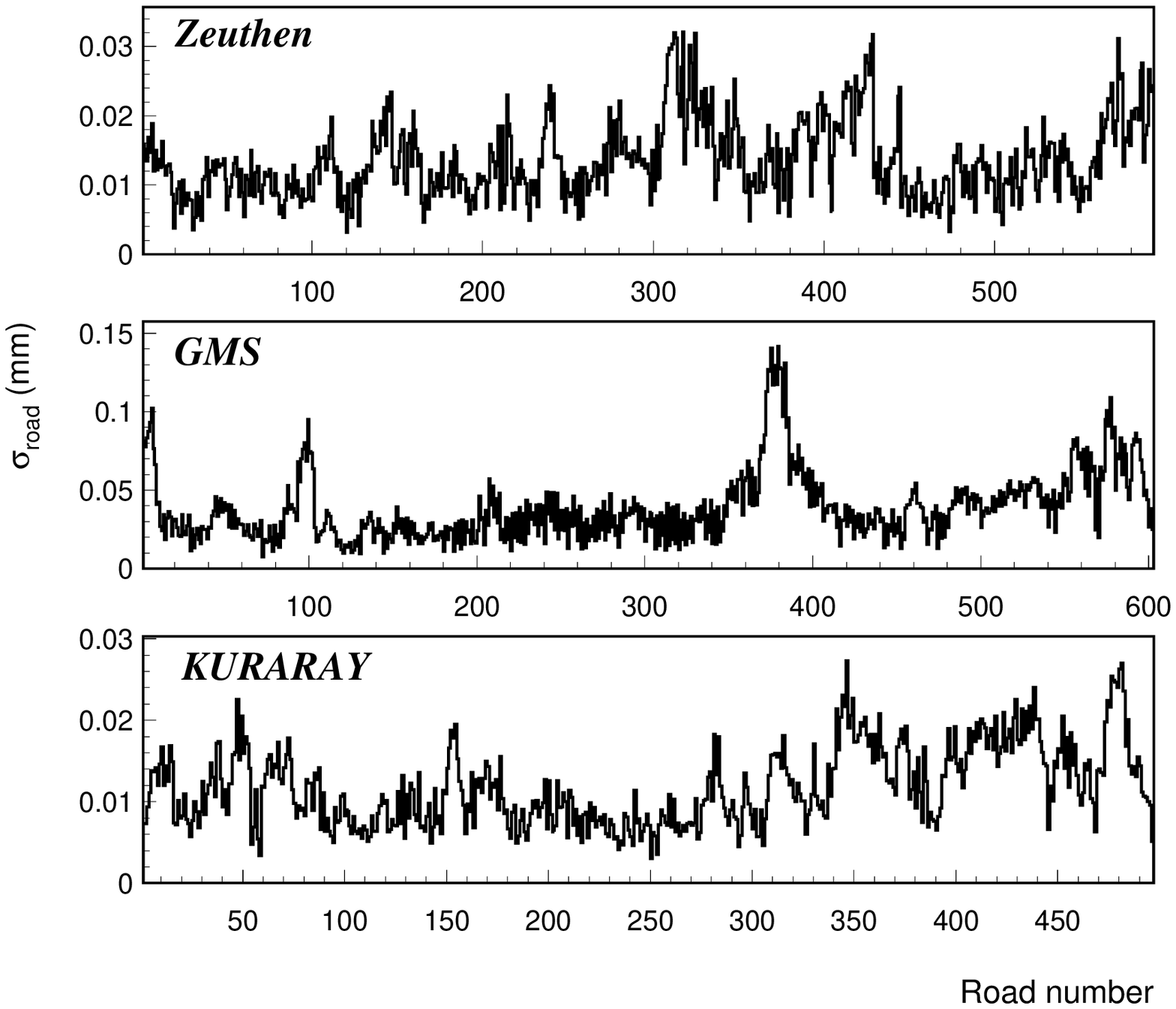,width=7cm}
\caption{Deviation of fibers from ideal position per fiber road
for three prototype detectors from Zeuthen, GMS and KURARAY.}
\label{bild6}
\end{minipage}
\end{figure}

\begin{figure}[b] 
\begin{minipage}[b]{7cm}
\epsfig{file=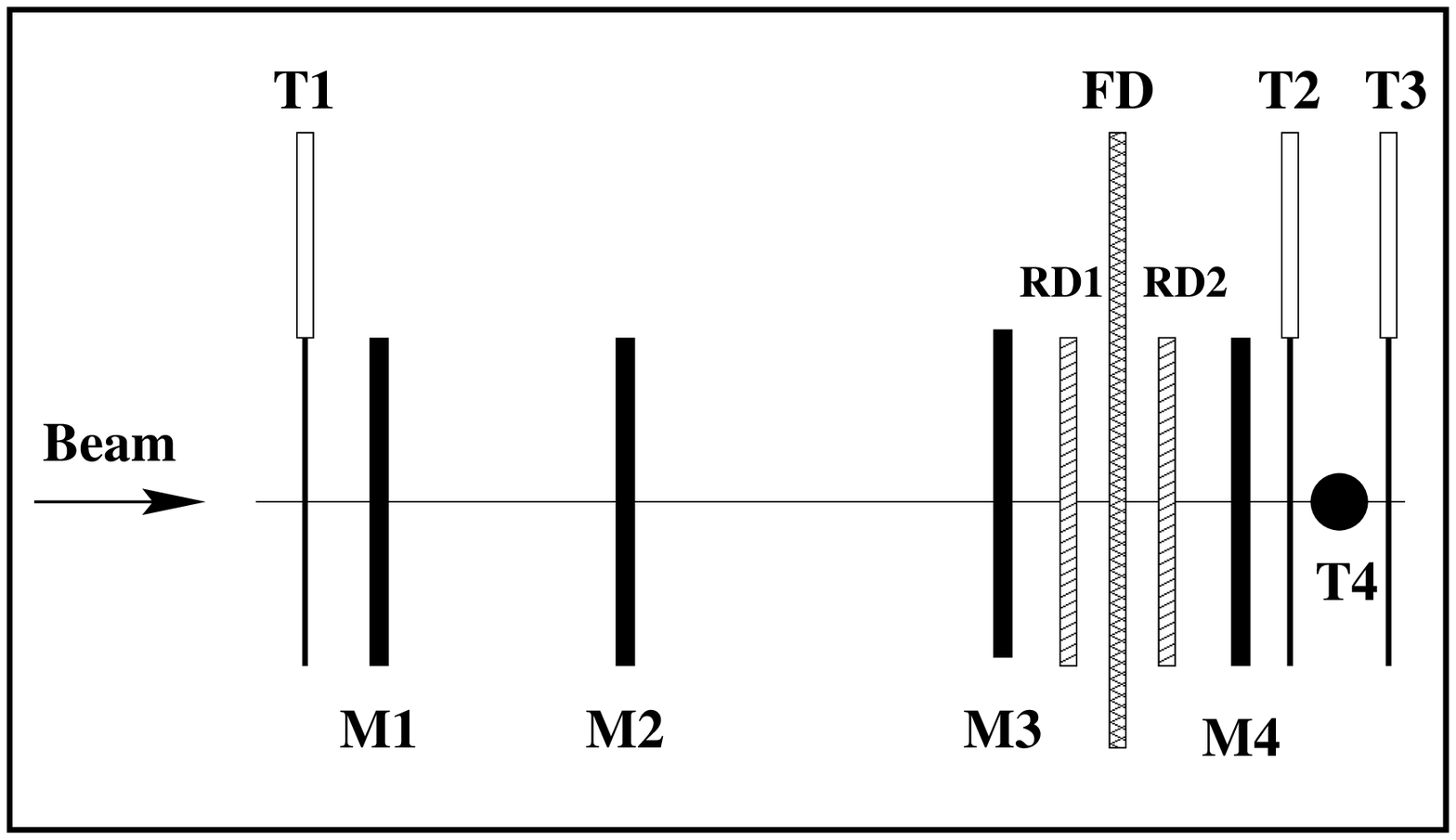,width=7cm}

\vspace{1.0cm}
\caption{Set up for testruns 1 and 2. M1-M4: silicon microstrip
 detectors, RD1 and RD2: fiber reference detectors, T1-T4:
 trigger paddels, FD: detector to be tested.}
\label{bild7}
\end{minipage}
\hfill
\begin{minipage}[b]{7cm}
\epsfig{file=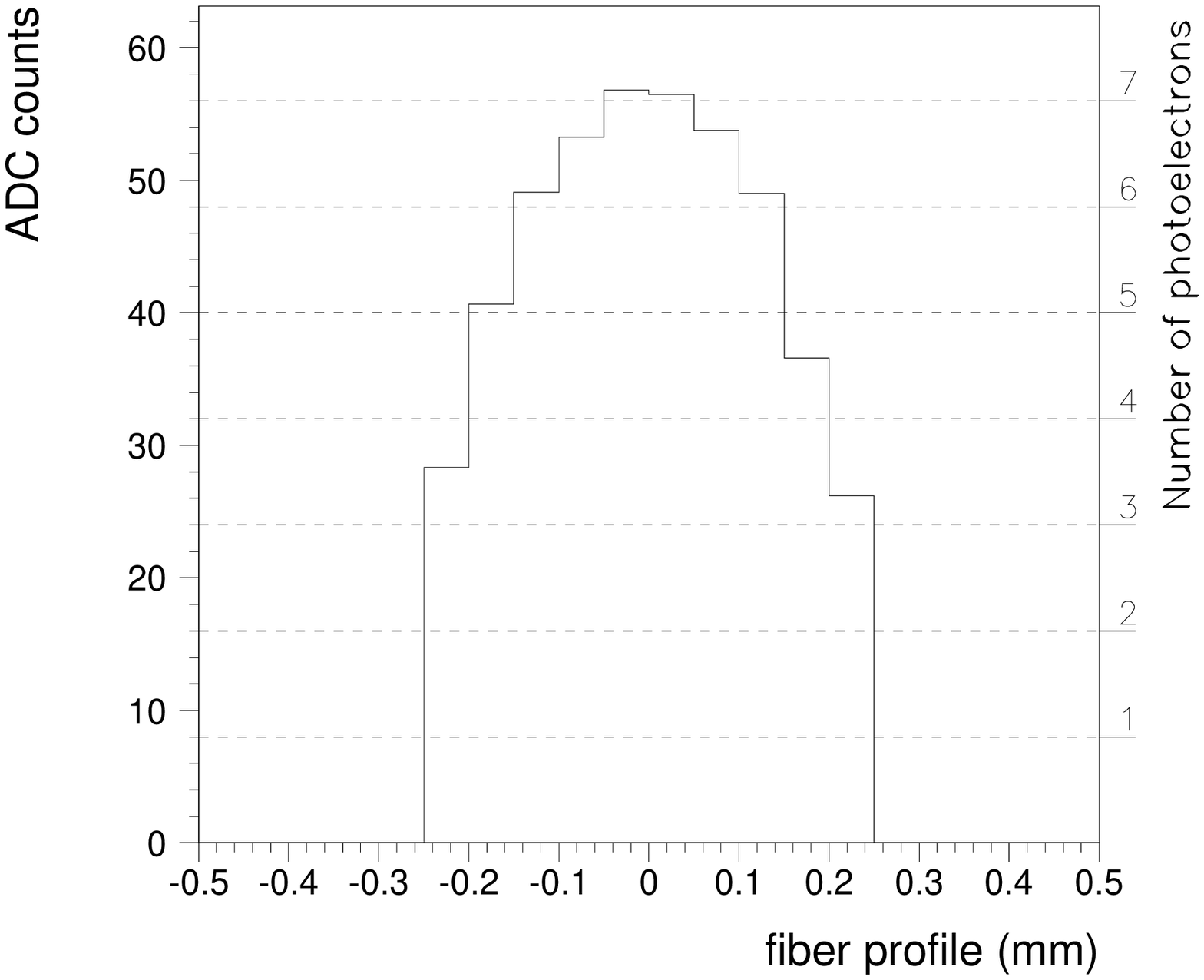,width=7cm}
\caption[]{Light output across a fiber road of seven 500 $\mu$m KURARAY 
fibers coupled to a 3 m long light guide of 1.7 mm diameter.}
\label{bild8}
\end{minipage}
\end{figure}

\clearpage

\newpage

\begin{figure}[t] 
\begin{center}
\begin{minipage}[t]{8cm}
\epsfig{file=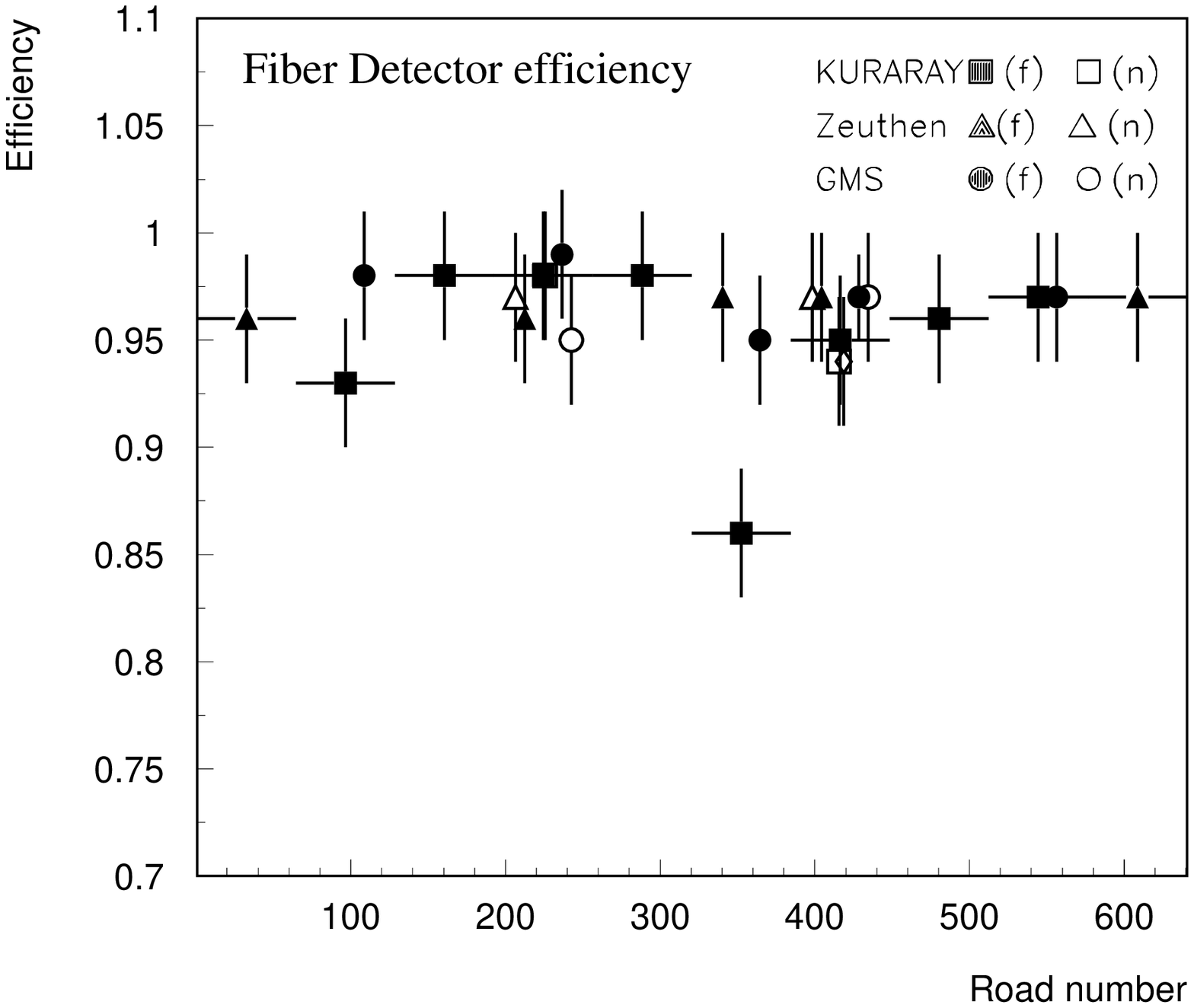,width=8cm}
\caption{Efficiency at different positions of three fiber
detector prototypes, averaged for 64 channels. Particles hit 5cm
from the near (n) or far (f) end of the ordered detector part.}
\label{bild9}
\end{minipage}
\end{center}
\hfill
\begin{center}
\begin{minipage}[b]{8cm}
\epsfig{file=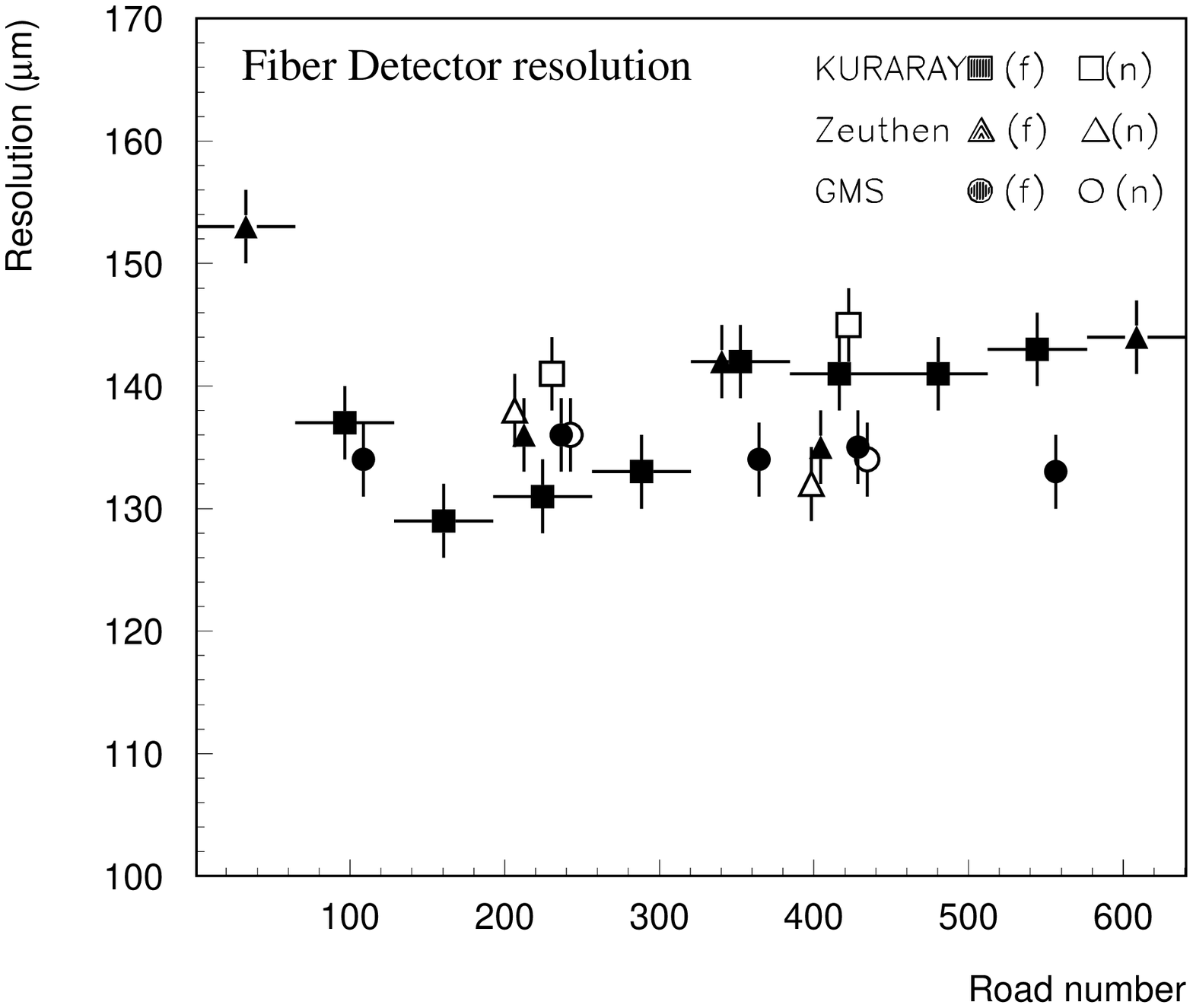,width=8cm}
\caption{Resolution at different positions of three fiber
 detector prototypes, averaged for 64 channels.  Particles hit 5cm
from the near (n) or far (f) end of the ordered detector part.} 
\label{bild10}
\end{minipage}
\end{center}
\end{figure}


\begin{references}
\bibitem{lit1}Ansorge, R., et al., {\it NIM} {\bf 265}, 33 (1988)
\bibitem{lit2}Annies, P., et al., {\it NIM} {\bf A367}, 367 (1995)
\bibitem{lit3}Bross, A.D., {\it Nucl. Phys. B (Proc.Suppl.)} {\bf 44},
        12 (1995)\\
       Adams,D.,et al., {\it Nucl. Phys. B (Proc.Suppl.)} {\bf 44}, 332 (1995)
\bibitem{lit4}B\"ahr, J., et a., {\it Proceedings of the 28th Intern. 
              Conf. on High Energy Physics, Warsaw, Poland, 1996, 
              eds. Z.Ajduk,A.K.Wroblewski} {\bf V. II}, p. 1759
\bibitem{lit5}Lohse, T., et al., {\it HERA-B Technical Proposal, 
              DESY-PRC} {\bf 94/02} (1994)
\bibitem{lit6}Ferro--Luzzi, M., et al., contribution presented by A.Gorin
              to this workshop
\bibitem{lit7}Yoshizawa, Y., contribution to this workshop
\bibitem{lit8}Aschenauer, E.C., et al., preprint {\it DESY} {\bf 97-174} (1997)
\bibitem{lit9}Nakano, T., et al., {\it Proceedings of the workshop SCIFI93,
              Notre Dame, USA, 1993, eds. A.Bross, R.Ruchti, M.Wayne}, p. 525
\bibitem{lit10} Eisele, F.,et al., private communication
\end{references}
\end{document}